\documentclass[10pt]{article}

\usepackage{amsmath}
\usepackage{amssymb}

\usepackage{graphicx}

\usepackage{cite}

\usepackage{color} 

\usepackage[labelfont=bf,labelsep=period,justification=raggedright]{caption}

\makeatletter
\renewcommand{\@biblabel}[1]{\quad#1.}
\makeatother

\date{}

\begin{document}

\begin{flushleft}
{\Large
\textbf{Robust Detection of Hierarchical Communities from {\em Escherichia coli} Gene Expression Data}
}
\\
Santiago Trevi{\~n}o III$^{1,2}$,
Yudong Sun$^{1,2}$,
Tim F.\ Cooper$^{3}$,
Kevin E.\ Bassler$^{1,2,\ast}$
\\
\bf{1} Department of Physics, University of Houston,
 Houston, Texas 77204-5005, USA
\\
\bf{2} Texas Center for Superconductivity, University of Houston,
 Houston, Texas 77204-5002, USA
\\
\bf{3} Department of Biology and Biochemistry, University of Houston,
 Houston, Texas 77204-5001, USA
\\
$\ast$ E-mail: bassler@uh.edu
\end{flushleft}

\section*{Abstract}

Determining the functional structure of biological networks is a central goal of systems biology.
One approach is to analyze gene expression data to infer a network of gene interactions on the 
basis of their correlated responses to environmental and genetic perturbations. 
The inferred network can then be analyzed to
identify functional communities. 
However, commonly used
algorithms can yield unreliable results due to experimental noise, algorithmic stochasticity, and the influence of arbitrarily 
chosen parameter values. 
Furthermore, the results obtained typically provide only a simplistic view of the network partitioned into
disjoint communities and provide no information of the relationship between communities.
Here, we present methods to robustly detect co-regulated and functionally enriched gene communities and 
demonstrate their application and validity for
  {\em Escherichia coli} gene expression data. 
Applying a recently developed community detection algorithm to the network of interactions 
identified with the context likelihood of relatedness (CLR) method, 
we show that a hierarchy of network communities can be identified.
These communities significantly enrich for 
gene ontology (GO) terms, consistent with them representing biologically meaningful groups. 
Further, analysis of the most significantly enriched communities identified several candidate new regulatory interactions. 
The robustness of our methods is demonstrated by
showing that a core set of functional communities is reliably found when
artificial noise, modeling experimental noise, 
is added to the data. 
We find that noise mainly acts 
conservatively, increasing the relatedness required for a network link to be reliably assigned
and decreasing the size of the core communities, 
rather than causing association of genes into new communities. 

\section*{Author Summary}

One of the fundamental themes in biology is the hierarchical organization of its constituents. At
higher levels of a hierarchy new properties emerge due to the complex interaction of constituents at 
lower levels.  This same organization is expected to be found in genetic regulatory networks. If so, determining this hierarchal structure would aid in understanding the properties and functional processes of the networks.
With the increasing availability of genetic expression data, developing methods to infer the underlying genetic regulatory network and detect functional communities within the network is an important goal of systems biology.
Unfortunately, noise in expression data creates variability in the inferred network and 
the stochastic nature of 
community detection 
creates variability in the functional communities detected with existing methods. Here, we present
methods for exploring the hierarchical organization of genetic regulatory networks that robustly detect core functional communities.
  We test the methods and demonstrate their validity, by applying them to  {\em Escherichia coli}
 genetic expression data,  finding a hierarchy of functionally relevant communities and then comparing those communities to the 
  known E. coli functional groups. 
  We then give examples of how our methods can be used to infer regulatory interactions
  between genes.

\section*{Introduction}

Gene regulation networks represent the set of regulatory interactions between all genes 
of an organism. These networks can contribute to our understanding of the development 
of organisms and how they integrate internal and external signals to coordinate gene expression 
responses~\cite{temporal, Davidson_gene_reg}. 
 Moreover, knowledge of gene regulation networks allows communities of 
closely interacting genes to be identified. Once identified,
such communities are an important resource for developing hypotheses for the function of uncharacterized genes and can provide insight into patterns of regulatory network 
evolution and function~\cite{Faith_Z, Ma,Cooper, Shi,module_from_expression, Bonnet}. Examining the relationships between communities can also reveal a hierarchical set of interactions, which is thought to be a fundamental organizing principle in many biological systems ~\cite{Beyer, hierarchicalorganization, Rev_Bio_network}. 
For all these reasons, determining gene regulation 
networks and their functional organization remains a major goal of systems biology.

The increasing availability of gene expression data has spurred development of a number of approaches that aim to determine the underlying structure of the transcriptional regulatory 
network~\cite{Davidson_gene_reg,module_from_expression,Beayesian_networks,Faith_Z,Probgraph,expressionprofiling,balazsi, two-way_clustering}. 
Most of these techniques fall into the broad categories of correlation-based methods, information-theoretic methods,
 Bayesian network predictions, or methods based on dynamical models. 
 These approaches generally infer regulatory links between the nodes (genes) of the network on the basis of the level of 
correlation in their transcriptional response to a series of environmental and genetic perturbations. 
The strength of the links is either weighted by the correlation value,
or is unweighted and the links are assumed to exist only if the
correlation exceeds a threshold value. Once the links are assigned,
the network becomes well defined.
However, variation in the application of each 
method can produce differences in the link weight between pairs of nodes.  
Additionally, if the threshold for placing links is varied even slightly there can be 
significant differences in the network structure inferred  from a given data set~\cite{inference}.   
Identification of groups of interacting node (gene) communities  poses an additional challenge. 
Communities can
be identified using computational methods 
developed in network science~\cite{communityreview}. 
These methods include
hierarchical clustering~\cite{Hierarchical_wen,Hierarchical_Eisen,Hierarchical_Weinstein}, clique based clustering~\cite{k-core,k-plex, overlapping, socialgroup}, core-pheriphery~\cite{19426456,cpdefine, cpdefine2},  
K means clustering~\cite{kmeans}, principal component analysis~\cite{PCA_cluste,SVD_expression}, label propagation~\cite{PhysRevE.76.036106, RePEc:spr:jeicoo:v:4:y:2009:i:2:p:221-235}, statistical mechanical approaches~\cite{PhysRevLett.76.3251, PhysRevLett.93.218701} , and modularity maximization methods~\cite{Qdef, Newman, PhysRevE.70.025101, PhysRevE.72.027104, FinalTuning}.
 Often these algorithms agglomerate or divide the nodes of a network
 into groups based on either the links of the network or the strength of the correlation value between
 pairs of nodes.  
 However certain algorithm parameters, such as the number of groups, are often required as user inputs and can become increasingly difficult 
 to predict
 as the size and complexity of the network grows. 
 In addition, 
there can be considerable variability in the community detection process due to approximations and stochastic elements of the 
computational algorithms.

Here, we present methods for determining the hierarchical organization of genetic regulatory networks and for detecting functional 
 communities of genes that are robust to variability in both gene expression data and community detection parameters.
We apply a recently developed community detection method~\cite{FinalTuning}
to regulation networks inferred from a 
compendium of {\em E. coli} expression profiles using the 
context likelihood of relatedness (CLR) algorithm~\cite{Faith_Z}.
This method uses the mutual information in the data sequence for
pairs of genes to construct a ``Z-score matrix'' that describes the relatedness
of each gene pair.
We then choose a threshold Z-score value and construct a network by
creating links between pairs of genes whose relatedness exceeds this
value. However, rather than choose one threshold value, we investigate the network
using a range of threshold values.
The combination of using the CLR method and varying the threshold value used to
create the network captures non-linearities inherent in the network structure. 
We identify communities using
a leading eigenvalue method with final tuning ~\cite{FinalTuning}. This method 
identifies communities by partitioning
the network so as to maximize its
modularity. The optimization algorithm used by this method, when applied to
a series of widely studied networks, produces the
partitioning with the largest modularity of any known fast algorithm
for networks up to a few
thousand nodes in size~\cite{FinalTuning}. 

As mentioned above, there is variability in the community detection process.
Indeed, numerous network partitions can give 
modularities close to the maximum and these partitions can be 
structurally diverse~\cite{Clauset}.  Rather than treat this property as a disadvantage, we 
use the stochasticity to find correlations between different runs of the community detection algorithm.
We consider a core community, as those nodes 
that are consistently assigned to the same community over multiple partitions of the network. 
This ensemble analysis of partitionings to find 
correlations between different sets of network partitions, combined with
varying the threshold value used to create a network, enables us to investigate  
relationships between communities at different threshold values. We define community relationships as hierarchical
if communities at a higher threshold value are contained within communities at a lower threshold value.
This method not only allows us to find the hierarchical organization of communities within the network,  but also to determine
if a network is, in fact, hierarchical -- a feature that is not forced upon the network by the method.

Comparisons of independent gene expression experiments often find considerable inter- and even intra-experiment variation, which can amplify stochastic aspects of the community detection process ~\cite{Chen, Baggerly, Duewer}.
While variation can be minimized by standardizing the platform and analysis pipeline used, the low-replication common to many gene expression studies, means that the variance of each individual gene expression estimate is typically quite high. To investigate the effects of experimental noise on our ability to assign genes to core communities, we constructed artificial data sets with various levels of experimental noise. 
At each noise value,  multiple runs of the community detection process are performed, allowing us to determine the sensitivity of core community structure to realistic levels of expression variation. 
We find that increasing the value of expression noise had a similar effect to increasing the relatedness cutoff value used to create the network. Noise decreases the size of the core communities, leaving only the most strongly related genes as consistent members, but does not tend to assign genes into new core communities. To test whether the communities predicted by our methods are biologically relevant, 
 we test whether they significantly enrich for gene ontology (GO) 
terms identified in {\em E. coli}. We find that, in many cases, there are statistically significant matches between a core community and GO term, indicating that communities are biologically relevant. Thus, the methods we present to investigate genetic regulatory networks and to determine the hierarchy of their functional communities 
appear robust to the variability in the community detection process  and to the existence of experimental noise.

\section*{Results}

\subsection*{Inferring gene interaction networks from expression data}

\begin{figure}[ht]
\begin{center}
\includegraphics[width=4.5in]{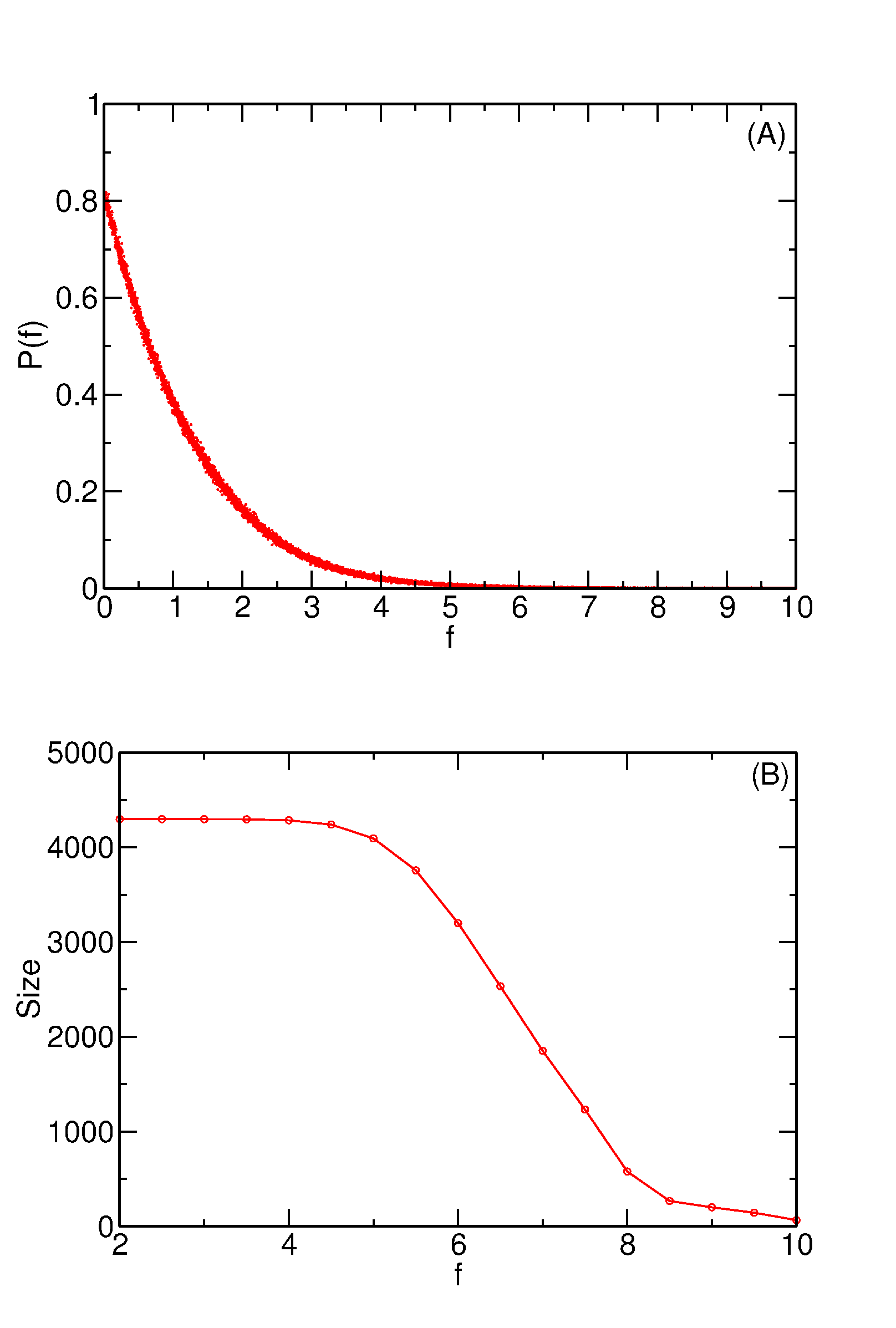} 
\end{center}
\caption{
{\bf Distribution of gene relatedness and network size in the {\em E. coli} CLR network.} (A) Probability distribution of relatedness values, $f$,  between pairs of genes in {\em E. coli} calculated using the CLR algorithm and the full ${\rm M^{3D}}$ dataset. (B) Size of the largest connected component for relatedness value, $f$. At small values of $f_{\min}$ the network is fully connected but begins to break up into multiple disconnected components at a critical value of approximately $f_{\min} = 4$.}
\label{fxydist}
\end{figure}

We used the CLR algorithm to infer direct and indirect regulatory interactions between {\em E. coli}  genes on the basis of the similarity of their expression response in 466 experiments in the Many Microbe Microarrays
Database (${\rm M^{3D}}$)~\cite{M3d_data}. 
The resulting CLR relatedness matrix can be used to define a network with weighted links between genes. In principle this network can be analyzed to find its community structure. However, doing so would not allow an exploration of hierarchical community organization. 
Instead, we apply a threshold value of relatedness, $f_{\min}$, above which a regulatory interaction is inferred. The result is an unweighted, undirected network where links between genes indicate regulatory correlations. Note that these correlations do not necessarily imply direct interactions.
A link may indicate indirect interactions, as may occur between two genes if they are both regulated by a third gene. In this way the CLR network differs from annotated regulatory networks (e.g., for {\em E. coli} RegulonDB~\cite{Gama-Castro04112010}) that include only direct regulatory links. The threshold value $f_{\min}$ that is chosen has considerable effect on the network that is created and on its community
structure. The distribution of relatedness value,  $f$ , of pairs of genes is shown in Fig.~\ref{fxydist}A. 
Clearly, increasing the cutoff value significantly reduces the number of links in the network.  At $f_{\min} = 2$ all 4,297 genes are in the largest connected component and therefore the network is fully connected (Fig~\ref{fxydist}B). At approximately $f_{\min}
= 4$, the inferred network begins to break up and at $f_{\min} = 6$, the size of the largest connected component is substantially reduced and a number of isolated components exist. 
Thus, $f_{\min} = 4$ is approximately the critical value at which the network remains largely intact as one connected network. 
 In the work below we consider networks inferred from  $f_{\min}$ values of 2, 4 and 6.  These values correspond to points on, and at either side of the critical  threshold value. A list of the links in the network  $f_{\min} = 4$  and $f_{\min} = 6$ is given in Dataset S1.

\subsection*{Identifying communities and their hierarchical organization}

We used a recently developed extension of the leading eigenvalue method to determine
the community structure of the inferred {\em E. coli} regulatory network~\cite{Newman}. This method
aims to identify a partitioning of nodes
into a disjoint set that maximizes network modularity.  
Modularity, $Q$, is defined as the fraction of links that connect nodes
in the same community minus the fraction expected if
the partitioning and the degree sequence of the network remains fixed, 
but the links are randomly distributed~\cite{Qdef}. This definition of modularity quantifies the intuitive notion that one expects there to be more links between nodes of the same community
than between nodes of different communities, adding the constraint 
that the number of links inside a community should be larger than one would expect by chance. The definition is normalized so that the maximum possible value of $Q$ is 1. The larger the value of the modularity found by a partitioning, the more ``modular" a network is. A completely nonmodular network would correspond to $Q = 0$.  The extension of the leading eigenvalue method that we use, known as final tuning, is an extra step in the algorithm, related to the so called Kernigan-Lin algorithm~\cite{kernighan_lin}, that removes systematic biases and produces the best results of any known fast modularity maximizing algorithm for networks of the size considered here. 

Community detection algorithms, including the one we use, contain stochastic elements that can cause different runs to give different partitionings. Indeed, partitionings of the same network can be structurally diverse, despite having similar modularity scores~\cite{Clauset}. Here, we exploit this property, by analyzing an ensemble of partitionings and measuring their correlations. This allows us to both find the pairs of genes that are most often grouped together and examine the family of community structures that can result from a modularity maximization.  

\begin{figure}[ht]
\begin{center}
    \includegraphics[width=5.5in]{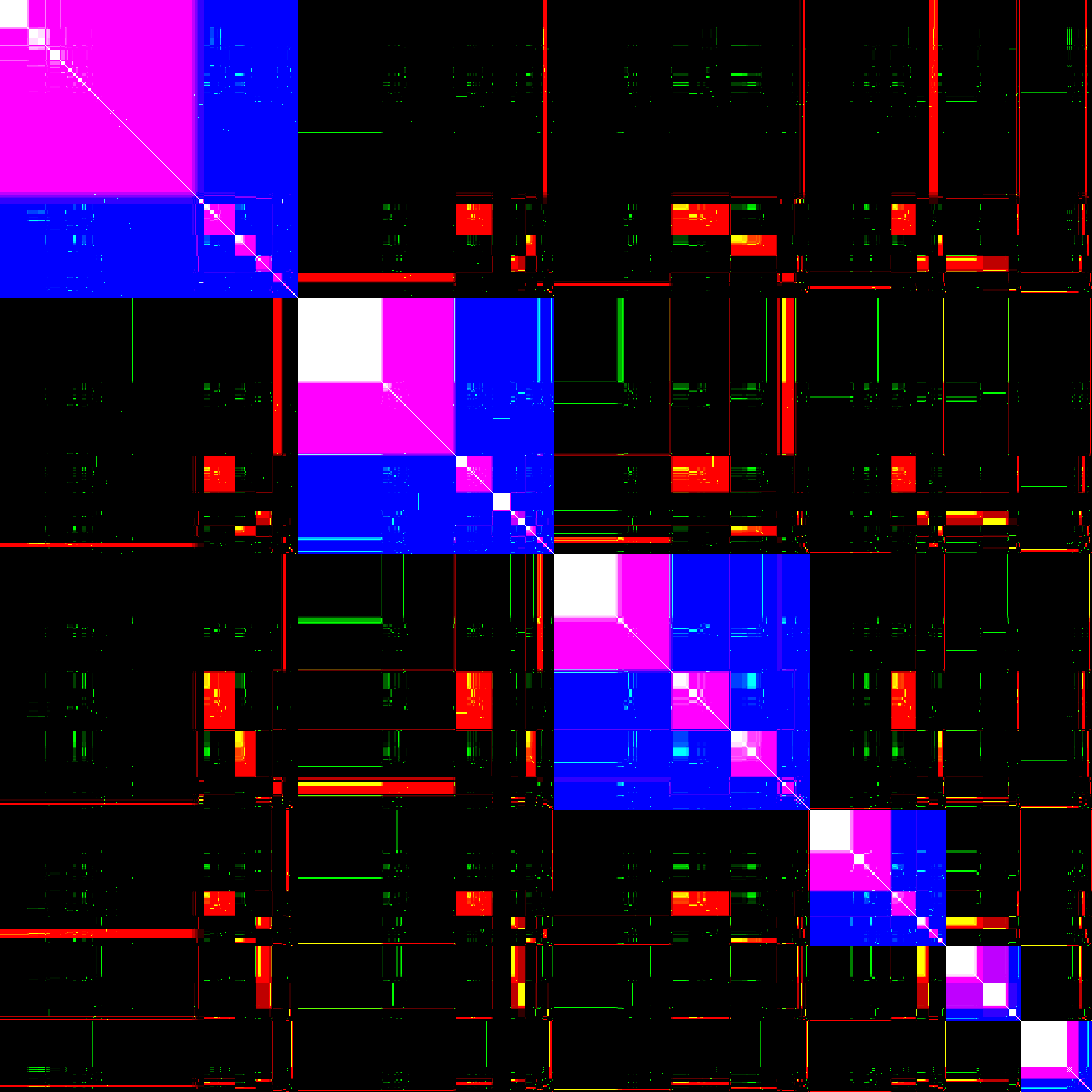}
    \caption{
    {\bf Correlation matrix showing community structure found in the {\em E. coli} network with relatedness threshold values $f_{min} = 2, 4,$ and $6$.}
    Genes are ordered in the same sequence
along the x and y axes beginning in the upper left corner, and this ordering is the same for all three relatedness values (gene order is given in SI).
The matrix element
in the position $(X,Y)$ is colored blue, red, or green if genes $X$ and $Y$ are
in the same community at threshold values 2, 4, or 6, respectively.
The density of the color indicates the strength of the correlation in the partitionings of the pair of genes. For example, considering the correlation between a pair of genes in the 10 replicate partitionings performed on the $f_{min} = 4$ network, dark and light red indicates that the pair of genes 
are always and rarely found to be in the same community, respectively. 
The red, green and blue colors corresponding to $f_{min} = 2$, 4 and 6 thresholds, respectively, are combined to indicate the correlations of each pair of genes at all three threshold values.  
Thus, the color of the matrix element
in the position $(X,Y)$ is white if genes $X$ and $Y$ are
in the same community at all three threshold values.
It is purple (yellow) if the two genes are
in the same community at thresholds 2 and 4 (4 and 6), but not at threshold 6 (2) and it is black if the two genes are not in the same community at any of the three
threshold values. A list of the order of genes is given in Dataset S2. A full size version with each pixel representing a distinct pair of genes is given in Fig. S1.
  }
    \label{ori_c2_c3}
\end{center}
\end{figure}

At a particular $f_{min}$ value, which defines a unique network,  we ran our community detection algorithm 10 times, generating a correlation matrix where each element represents the proportion of times gene $X$ and gene $Y$ are found in the same community. 
We define sets of genes that are always found in the same community as a ``core community". 
We performed this procedure for $f_{min} = 2$, 4, and 6, which, as discussed above, give networks that are supercritical, critical, and subcritical, respectively.
Combining the three resulting correlation matrices generates a visual representation of the overall structure of the network (Fig.~\ref{ori_c2_c3}). 
A list of genes in each core community for $f_{min} = 2, 4,$ and $6$ is given in Dataset S3.

We find substantial differences in the community structure of the networks inferred at different $f_{\min}$ values (Fig.~\ref{ori_c2_c3}).
As $f_{\min}$ is increased, links that connect weakly related genes are removed from the
network, which can cause genes to switch communities, and communities to
merge or divide. Analysis of these changes lead to two conclusions.
 First, there is a basic community structure that is robustly determined such that many pairs of genes remain in the
same community at all three $f_{\min}$ values, indicated by the block diagonal white elements. That
is, there is a basic community structure that is invariant
with respect to adding or subtracting links between weakly related
genes. Second, community structure is hierarchical. 
To see this, note that at  $f_{min} = 2$ the community structure consists of six large communities, indicated by the blue blocks, while at higher values it begins to break up into smaller communities. More importantly, the relationship between communities at different $f_{\min}$ values indicates that the structure of the network is largely hierarchical.  A hierarchical structure is revealed when a community breaks up into subcommunities as $f_{\min}$  increases. 
If the  {\em E. coli}  regulatory network was completely hierarchical, we would see only block diagonal elements consisting of large blue blocks that break up into purple then white sub-blocks as $f_{\min}$ is increased. Communities at one value of $f_{\min}$ that are subcommunities of the same community at a smaller $f_{\min}$  value are therefore hierarchically closer to each other than ones that remain in different communities at the smaller $f_{\min}$  value. 
Figure~\ref{ori_c2_c3} indicates that the inferred {\em E. coli} regulatory network has a largely but not completely hierarchical structure. This is apparent from the large fraction of the blue blocks ($f_{\min} = 2$ communities) that contain on diagonal purple and white blocks ($f_{\min} = 4$ and 6, respectively). However, there are some red off diagonal blocks that indicate a non-hierarchical ordering as $f_{\min}$  is increased from 2 to 4.   Furthermore, although the purple  $f_{\min} =  4$ blocks largely break up into white blocks as $f_{\min}$  is increased to 6, there are some off diagonal cyan and green blocks that indicate non-hierarchical ordering.  About $68 \%$ of the core community matrix elements at $f_{\min} = 4$ were hierarchically in core communities at $f_{\min} = 2$,  and about 80\% of the core community matrix elements at $f_{\min} = 6$ were hierarchically in core communities at $f_{\min} = 4$.
The organization of genes shown in this plot, is given in Dataset S4, where the, blue, purple, and white module membership of each gene is listed.

At  $f_{\min} = 2$ there are only six communities,
while at  $f_{\min} = 6$ there is a mode of 965 communities with the largest consisting of 417 genes.
This is consistent with the finding that at small values of
$f_{\min}$ the network is fully connected, while at large values the network
breaks up into a large number of small disconnected parts.  
At intermediate values of the threshold, where the network begins
to break up, the community structure is complex,
consisting of a broad distribution of different sized communities. Interestingly, as $f_{\min}$ increases so does the value of the maximum
modularity found, $Q_{\max}$. At $f_{\min} = 2$,
$Q_{\max}\approx 0.37$ indicating that the network structure is not particularly
modular, while at  $f_{\min} = 6$,
$Q_{\max}\approx 0.85$ indicating that the network structure is highly modular.

\subsection*{Community structure is robust to experimental noise}

Given the relatively high experimental variation and low replication typical of gene expression measurements, it is of practical interest to determine whether inferred community structure is robust to this source of noise. To address this question,
we consider a restricted set of the gene expression data comprising the 152 experiments
present in the ${\rm M^{3D}}$ database that were repeated at least three times.
 For each of these experiments,
a mean value $m(X)$ and a standard error $\sigma(X)$ for the expression
level of each gene $X$ is calculated. These values are used to generate artificial datasets with a variable level of noise, $c$. 
For a value of $c=1$, the artificial data sets have noise levels consistent with the experimental data. 
For larger (smaller) values of $c$, the artificial datasets have more (less) variability in the expression of each gene, than the experimental data. 
For each of a number of values of $c$, ranging from 0 to 4, 20 artificial data sets are produced.
Crucially, these data sets considered each gene and experiment independently, thereby preserving any inherent differences between different gene's expression variability. 

\begin{figure}[ht]
\begin{center}
\includegraphics[width = 4.0in]{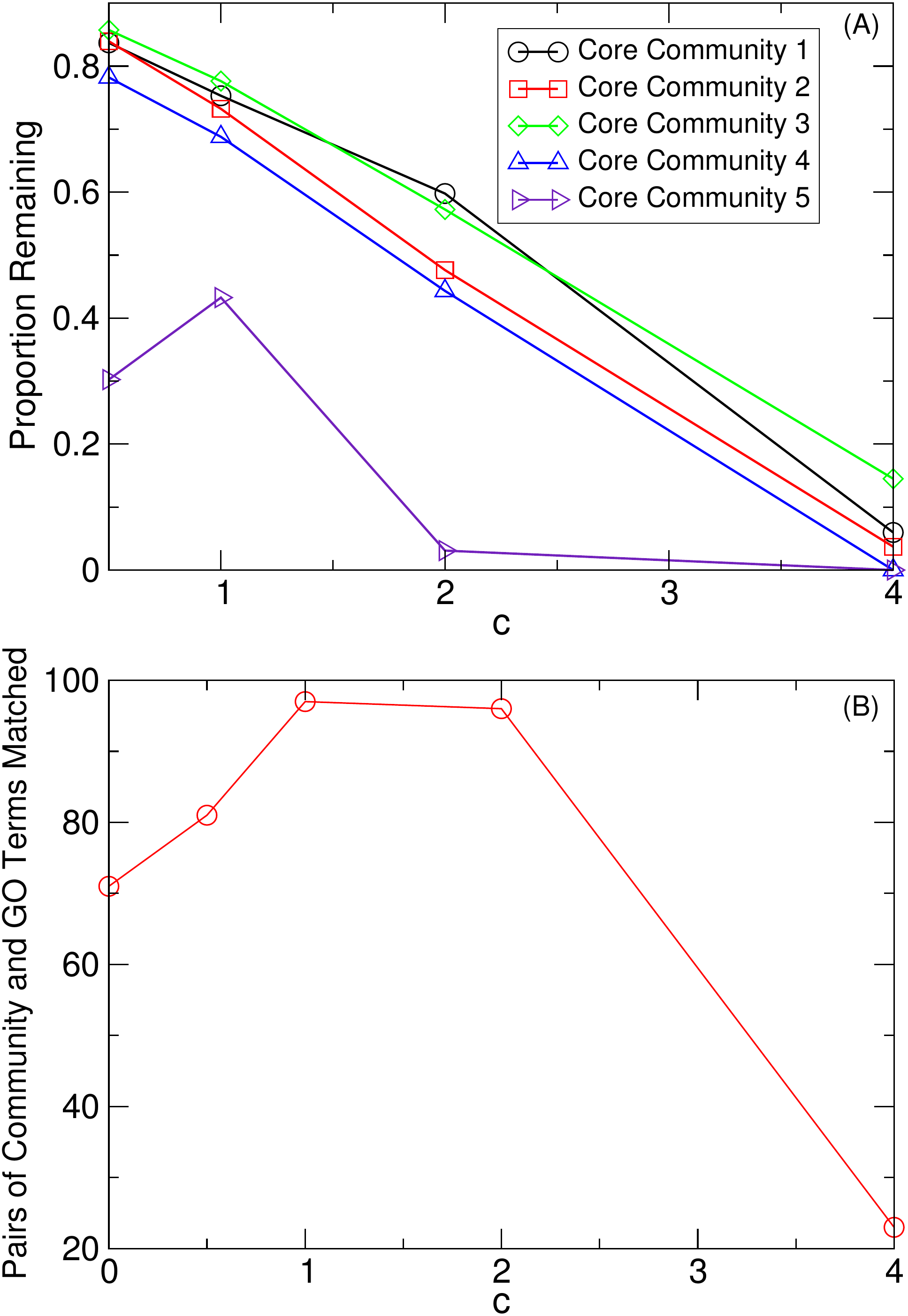}
\caption{
{\bf The effect of noise on core community structure and GO term enrichment.} 
(A) Proportion of $c=0$ core community nodes that remain in a core community.
(B) The number of significant GO term enrichments as a function of noise level $c$ for networks constructed with $f_{min} = 2$.
  If a GO term is enriched by more than one community, each enrichment is counted separately.}
 \label{propretained}
\end{center}
\end{figure}

For each noisy data set, we used the CLR algorithm to infer a regulation network at an
$f_{\min}$ value of 2, and the community structure was
determined with the methods described above.
For each dataset, 10 different community partitionings were obtained, giving a total of 200 partititonings for each value of $c$.
Figure~\ref{core_communities} shows a series of correlation matrix plots for the community structure found for the
partitioning ensembles for $c=0,0.5,1,2$ and $4$. The degree of noise clearly has a major impact on community structure.
Nevertheless, except at $c=4$, there exist robustly determined core communities.
In addition this analysis revealed two important results. First, as the noise level $c$ increased, a large proportion of the genes in a core community
are partitioned into sub communities but genes rarely switch out of their $c = 0$ core communities. 
  This is similar to what happens when the threshold value for creating the network was increased (Fig.~\ref{ori_c2_c3} ).
  Second, with one exception, the number of nodes included in each core community decreased as $c$ was increased (Fig.~\ref{propretained}A). We conclude that noise acts mainly conservatively, decreasing the size of core communities, rather than causing association of genes into new communities.

\subsection*{Communities enrich for functionally related genes}

We have thus far demonstrated that our computational methods can robustly identify a community structure in the {\em E.\ coli} regulatory network. An important remaining question is
whether this structure is biologically relevant. To test this, we first examined the simple expectation that genes in the same operon, and that therefore share at least one promoter control region, will tend to group together in the same community. Even using the very stringent requirement that all genes within an operon be in the same community and not accounting for the presence of secondary promoters that are internal to the operon and might act to decouple operon regulation, we find that genes within an operon are much more likely to group together that expected by chance (Permutation test, {\em p}  $<  0.001$)(Fig. S3). For example, given the number and size of communities found at $f_{\min} = 6$, approximately 1\% of operons remain together if individual genes are assigned to communities randomly, compared to  $>45\%$ in the community assignments determined by the final tuning algorithm.

Next we asked whether the  
 community structure inferred by our method
groups genes with similar biological functions. 
To do this, we tested whether the identified communities significantly enrich 
for any of the gene ontology (GO) terms identified in {\em E.\ coli} ~\cite{GO,EcoCyc,EcoliHub}. 
(Note only core communities larger than 10 were considered because the method we use to partition the network will not accurately identify small communities~\cite{resolution}.) We found 147, 239 and 288 statistically significant matches between core communities and GO terms for communities identified at $f_{\min}$ values of $2, 4$ and $6$, respectively.  
Table~\ref{compare} details these results for the 25 most 
enriched relationships found at 
$f_{\min} = 4$ (complete tables of GO enrichments at $f_{\min}$ values of 2, 4, and 6, 
and GO terms used are given in 
Dataset S7 and Dataset S8, respectively). Note that many genes are described by multiple GO terms, e.g., the gene $flgM$ is a member of all terms in the GO hierarchy: 'flagellin-based flagellum basal body, rod' $\rightarrow$ 'flagellin-based flagellum' $\rightarrow$ 'flagellum' so not all enrichments are independent. Nevertheless, our network partitioning results in communities that significantly enrich for many GO terms, suggesting that the gene groupings are biologically meaningful. 

Figure~\ref{propretained}B shows the number of statistically significant GO term enrichments as a function of noise level, $c$.  Interestingly, enrichment peaks at a noise level of $c=1$, which corresponds to the artificial data with noise level consistent with that of the experimental data. 
This is presumably due to the fact that the mean expression values found from the experimental data are estimates, so that a noise value of $c=0$ will give a precise, but not necessarily accurate estimate of gene expression.
 As discussed above, increasing the noise in the artificial datasets causes the size of the core communities to decrease.  Interestingly, the $c=0$ core community  that 
dissolves the quickest, core community 5 (numbered beginning in the upper left hand corner of Fig.~\ref{core_communities}A),  
contributes only one significant GO term enrichment at $c=0$ (full details in Dataset S9). Finally, we note that there are some differences in the identity of core communities when the restricted set of 152 experiments is compared to those generated using the full experimental data (at $f_{\min} = 2$).
Nevertheless, as mentioned in Ref.~\cite{Faith_Z}, the CLR algorithm can produce nearly equivalent results as the full data set when a small, yet diverse set of expression profiles is chosen. 
This fact highlights  the importance of judiciously choosing experimental conditions when the data set is small. 

\subsection*{Inferring candidate regulatory interactions}

Partitioning of regulatory networks into communities of genes with similar responses to genetic and environmental perturbations can be used to identify candidate new regulatory interactions between genes.  To this end, we consider the communities that most significantly enriched for a GO Term at  $f_{\min} = 4$  and  $f_{\min} = 6$, and compare the relatedness network among the genes within each community to the subnetwork of known regulatory interactions involving these genes presented in RegulonDB. We stress, however, that what follows are simply two examples. Our results, given in the supporting information, contain a wealth of other gene communities whose interactions can be analyzed in a similar manner.

The community with the most significant GO term enrichment at $f_{\min} = 4$ contains 72 genes, including all 24 genes in the GO term for bacterial-type flagellum (Table S1). Because of their co-regulation, the remaining 48 genes in this community are implicated as having some relevance for the development, function or control of the  {\em E. coli} flagellum. Indeed, of these genes, many have recognized roles in environmental sensing and signal transduction, functions that are physiologically upstream of flagellum control. An additional 11 genes in the community do not have any annotated function, but two of them, {\em ycgR} and {\em yhjH}, contain domains that are consistent with flagellum related activity and five of them ({\em yjdA} {\em yjdZ} {\em ynjH} {\em ycgR} and {\em yhjH}) are annotated as being regulated by at least one of the two characterized regulators present in the community ({\em flhDC} and the flagellum sigma factor, {\em fliA})~\cite{Riley, Gama-Castro04112010}. One further unannotated gene, {\em ymdA}, is connected to {\em flhDC} only in the CLR network, and is therefore a candidate for being connected to flagellum regulation as well as having a role in flagellum function. The pattern of connections in this community also serves to highlight the difference between the RegulonDB (direct regulatory links) and CLR (co-regulation) networks. We identify ten operons that interact with FlhDC in the CLR but not the RegulonDB network. These interactions might represent previously unknown direct interactions, but are probably best explained as indirect interactions mediated through their direct regulation by FliA, which is regulated by FlhDC (Fig. S4).

\begin{figure}
\begin{center}
    \includegraphics[width=7.0in]{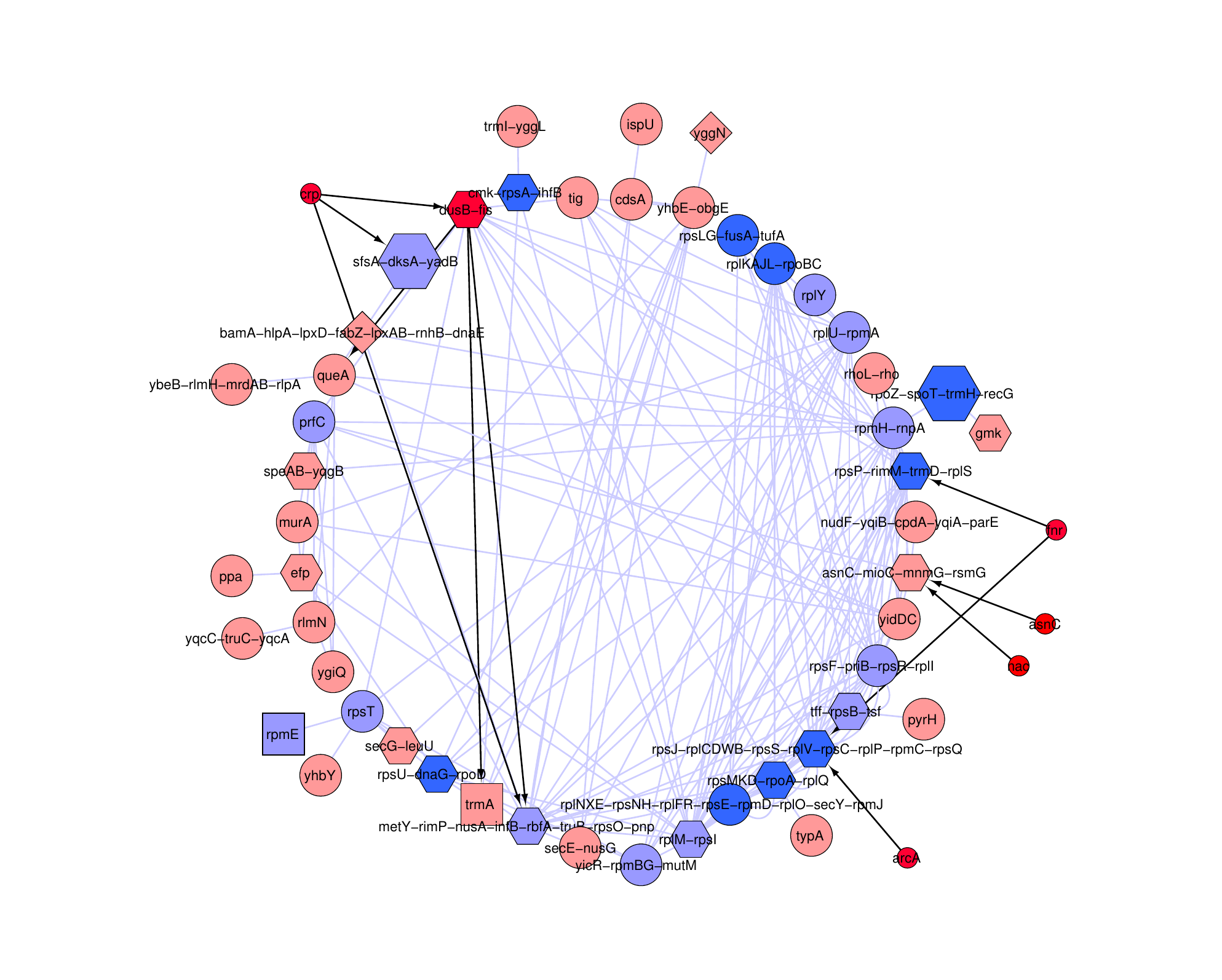}
    \caption{{\bf Links connecting operons in the $f_{\min} = 6$ community that enriches for genes involved in ribosome structure.}
    CLR links are in light blue, RegulonDB links are in black. Small symbols are genes that are not in the community, but are regulators of genes that are in the community and are therefore candidates for mediating indirect interactions between community genes. Symbol shape and color indicate attributes as follows: red, transcription factors; dark blue, ppGpp regulated promoter by direct assay~\cite{Lemke}; light blue, ppGpp regulated translation related promoter by microarray~\cite{MMI:MMI6229}; pink, other; hexagon, $\sigma70$ promoter; diamond, $\sigma$24 promoter; square, $\sigma$32 promoter; circle, unknown sigma factor. Note that very few interactions observed in the CLR network can be explained by the direct interactions annotated in RegulonDB. The high proportion of ppGpp sensitive promoters among operons contained in the community suggests this molecule as a good candidate for regulating the remaining interactions.  The network layout was determined by the circular layout option in Cytoscape 2.8.1, no particular significance should be attached to operons being outside the main circle. }
    \label{fmin6module}
\end{center}
\end{figure}

At $f_{\min} = 6$ the community with the most significant functional enrichment contains 107 genes, including 51 of 56 genes annotated as being structural components of the ribosome (Table ~\ref{OVERLAPf6}). This very significant enrichment suggests that the 15 genes present in the community that do not have any annotated function might also be involved in translational processes. The most striking aspect of this community, however, is that it contains only one recognized regulator, {\em fis}, which, as annotated in the regulonDB database, is involved in only a very small fraction of the inferred regulatory interactions (Fig.~\ref{fmin6module}). Moreover, no recognized transcription factor serves to indirectly connect regulation of more than three of the community operons and no sigma factor is unique to this community. These observations suggest the presence of some other regulatory factor that is in common to some or all of the genes in the community. One candidate for this factor is ppGpp, a small molecule which, in association with DskA, is known to affect regulation of many ribosome associated genes by decreasing the stability of the RNA polymerase open complex~\cite{Barker2001673}. 
Indeed, a recent study directly examined the effect of ppGpp on nine of the 51 primary promoters present in the community. In all cases, ppGpp was shown to affect promoter activity in at least one of the tested conditions and a comparison of global gene expression profiles of bacteria that differed in ppGpp levels, found that a further twelve promoters in the community differed in expression by at least 2-fold in response to ppGpp~\cite{Lemke, MMI:MMI6229}. Together, these results suggest the remaining 30 promoters in the community as candidates to also be affected by ppGpp.

\section*{Discussion}

We present unsupervised methods for determining communities of co-regulated genes and their hierarchical organization
based on expression data profiles collected under a variety of environmental and genetic perturbations.
Our methods combine the CLR algorithm and a tunable threshold value to infer the underlying regulatory network. We then use a statistical
ensemble analysis of the network partitionings that result from a
recently developed community detection algorithm to determine the network's community structure. Applying our method to {\em E.\ coli} expression
data we obtain three key results. i). Regulatory communities in {\em E.\ coli} are largely hierarchical 
so that the effect of increasing (decreasing) the $f_{\min}$ threshold is largely simply to split (combine) the communities found. 
ii) The structure of the inferred regulatory network is robust to relatively high experimental noise. iii) Regulatory communities significantly enrich for functionally related gene groupings. We discuss these findings in turn.

The technique we use applies a threshold to determine whether mutual information between the expression responses of two genes is sufficient to infer a connecting regulatory link. We find that the value of this threshold influences the size and unity of the inferred network. However, the network structure is relatively invariant to the addition or removal of links between more weakly related genes. We note that there at least two broad mechanisms that might cause genes to be weakly connected in our network. First, the relevant molecular interactions may exert weak expression control on the regulated gene. Second, the regulatory interactions might be environmentally dependent, being active in only a subset of the experimental conditions. Comparison of communities present in regulatory networks obtained at increasingly stringent thresholds indicates that the regulatory network is largely hierarchical such that large communities present in the low threshold network tended to split into smaller sub-groups of strongly related genes as the threshold was increased. By contrast, increasing the threshold causes relatively few genes to associate in new communities that were not subsets of the original communities. 

Relatively high experimental noise is of considerable concern in analysis of gene expression data. Indeed, even small differences in preparation and sample growth conditions, or in the exact platform and analysis procedure used, can manifest as substantial differences in gene expression estimates~\cite{Chen, Baggerly, Duewer,Irizarry}.
 To address the influence of  experimental noise on our ability identify regulatory interactions and communities, we generate datasets with different noise levels, calculated independently across experiments and genes. Comparing communities identified in networks inferred from these data sets, we find that not only are our predictions for the functional communities robust against noise up to double that seen in the original empirical dataset, but that the effects of 
experimental noise are mainly conservative. That is, experimental noise reduces the size of core regulatory communities but does not tend to create new communities.

For the purpose of identifying functional communities in a biological network,
we find that it is useful to study the community structure of different networks constructed
with a range of relatedness threshold values. At large threshold values, the nodes
in each of the small disconnected pieces are highly related. These small groups provide the 
most statistically significant enrichments for GO terms and thus best identify biologically relevant communities. However, as the threshold value used to construct
a network is reduced, the community sizes tend to increase. These
enlarged communities include other nodes that may also be relevantly
related to the core communities found at higher threshold values. Because of these competing considerations,
if only one threshold value is to be chosen for which to make biological comparisons, we suggest that the critical threshold value
should be used, which for {\em E.\ coli} is approximately $f_{\min} = 4$. Choosing the critical value
will not only balance the above two considerations, but as discussed earlier, also gives the most statistically complex 
distribution of community structure.

The usefulness of our methods are multifold. First, the functional community predictions of the methods can be used to refine existing knowledge of the
functional relationships of genes in well known organisms such as {\em E.\ coli}. 
That is, the overlap of the core communities we find to the {\em E.\ coli} GO Terms is not exact, suggesting
that the additional genes in our core communities that enrich a particular GO term may themselves be candidates for genes that
should be included in that term of the gene ontology. In this way, the predictions of our method can be used to suggest
new experiments to refine our understanding of the {\em E.\ coli} regulatory system. 
We have explicitly demonstrated how this can be done by analyzing two of the communities found with our methods that significantly enrich 
GO terms and predicting previously unknown regulatory interactions. 
Furthermore our methods
can readily be applied to expression data for other, less well studied,
organisms, and to other types of biological data, to identify
functional communities in their networks. 
The predictions from our
unsupervised methods will be particularly useful, for making  initial approximate predictions for the functional communities and their organization of less well known organisms. Additionally, it should be noted that we have applied our methods to expression data based on an arbitrary variety of experimental and genetic perturbations. However, the methods could instead be applied to more targeted sets of expression data. For example, data based on particular types of environmental perturbations, particular types of genetic knockouts, with cells in a particular stage of the cell cycle, or with cells in a particular developmental stage of a multi-cellular organism. By examining more targeted data of these sorts, the dynamics of particular functional communities can be explored. 

\section*{Methods}

\subsection*{The expression data analyzed}

We analyze {\em E. coli} expression data downloaded from the Many Microbe Microarrays
Database (${\rm M^{3D}}$) version 4, build 5~\cite{M3d_data}. 
This build consists of a compendium of expression profiles from 730 different experiments
reporting expression of 4,298
{\em E. coli} MG1655 genes.
These experiments report the effect on gene expression of 380 different perturbations, of which
152 were repeated at least three times. Experiments include environmental perturbations 
such as PH levels, growth phase, presence of antibiotics, temperature,
growth media and oxygen concentration, as well as genetic perturbations.
For each gene the data from the various experiments were
normalized to account for varying detection efficiencies
and differences in labeling. The values then reported
are the $\log_2$ of the normalized expression intensity.

\subsection*{The context likelihood of relatedness method}

To identify interactions between genes we apply the context likelihood of relatedness (CLR) algorithm~\cite{Faith_Z}. 
Generally, network inference is difficult because of bias from uneven condition sampling, upstream regulation,
and inter-laboratory variations in microarray results. The CLR algorithm attempts to
mitigate these difficulties by increasing the contrast between the physical interactions and the indirect relationships by taking the context
of each interaction and relationship into account. 
Links are assigned based on the mutual information in gene expression patterns, which, unlike simple correlation methods, can accommodate non-linear relationships between pair-wise gene expression patterns. 
Although some other algorithms offer higher precision in terms of recovering known regulatory links~\cite{Zare}, 
CLR is attractive for allowing identification of indirect links that might serve to strengthen relationships between genes within co-regulated communities. We note, however, two limitations of networks derived from the underlying data set and CLR approach we use. First, the expression experiments are not considered as time series, which could give information as to the direction of regulatory interactions~\cite{Yeung29112011}. Second, we do not consider combinatorial regulatory interactions, for example, in which two or more regulator genes must be active to regulate a target gene.

 Our implementation of the CLR algorithm begins by calculating the
mutual information in the expression data for each pair of genes. This is done by treating the data for each gene as a
discrete random variable, so that every pair of genes $X$ and $Y$
is assumed to have expression levels $x_i$ and $y_i$ for
each experiment $i=1,2,3,\ldots$.
The mutual information $I(X,Y)$ in the expression of $X$ and $Y$ is
\begin{equation}
I(X,Y)
=
\sum_{i,j} p(x_i,y_j) \; \log \frac{p(x_i,y_j)}{p(x_i) \; p(x_j)}
\end{equation}
where
$p(x_i)$ and $p(y_j)$
are the
marginal probability distributions
that the expression level of $X$ is $x_i$ and
of $Y$ is $y_j$, respectively,
and
$p(x_i,y_j)$
is the  joint probability distribution
that, simultaneously, the expression levels of $X$ and $Y$ are $x_i$ and
$y_j$, respectively.
These discrete probability distributions are calculated from the
continuous expression data using B-spline smoothing and
discretization. Rather than assign an expression value to one bin, as in classical binning,
the B-spline functions allow an expression value to be assigned to multiple bins to account for 
fluctuations in biological and measurement noise. This is
sometimes referred to as  ``fuzzy binning"~\cite{B-spline}.
For $N$ genes, this calculation results in an $N \times N$ symmetric
matrix of mutual information values.
Here, to calculate the probability distributions for {\em E. coli} we use 10 discrete bins and a third-order B-spline
function. The results do vary slightly if the number of bins used or the order of the B-spline function is changed. 
However, the results vary slowly with these parameters and do not change any of our principle conclusions.

Mutual information between a gene pair can be due to random background effects,
or a regulatory relationship. To distinguish the relevant mutual information from its background, the CLR algorithm compares each mutual information value 
$I(X,Y)$, to the distribution of the mutual information values between gene $X$ and all other genes $\{I(X,Y) ;\forall Y  \}$, and separately, to the distribution of the mutual information
values between gene $Y$ and all other genes $\{I(X,Y) ;\forall X  \}$. The distributions are assumed to be normal and a Z-score value, $Z_x$ and $Z_y$, is assigned to $I(X,Y)$ for distribution $X$ and $Y$, respectively.  The Z-score value of I(X,Y) compared to a normal distribution $i$, with a mean $\mu_i$ and standard deviation $\sigma_i$, is given by

\begin{equation}
Z_i
=
\frac{I(X,Y)-\mu_i}{\sigma_i} \; .
\end{equation}
Any value of $Z_x$ or $Z_y$  less than zero is set to zero. Finally, the relatedness value between gene $X$
and gene $Y$ is defined as 
\begin{equation}
f(X,Y)=\sqrt{Z_X^2+Z_Y^2}.
\label{f}
\end{equation}
For $N$ genes, this calculation results in an $N \times N$ symmetric
matrix of relatedness values.

Once this matrix of relatedness values is calculated, we infer a network of regulatory interactions by placing links between
every pair of genes whose relatedness value exceeds some threshold, $f_{\min}$. For a given $f_{\min}$ value, this procedure results in a defined interaction network. A list of the links in the network  $f_{\min} = 4$  and $f_{\min} = 6$ is given in Dataset S1.

\subsection*{Network community detection methods}

There are a number of different methods that can be used to determine
the community structure of a given complex network~\cite{Hierarchical_wen,Hierarchical_Eisen,Hierarchical_Weinstein,kmeans,PCA_cluste,SVD_expression}.
Here we use a method that aims to find a partitioning of nodes
of the network into disjoint sets that maximizes the modularity of
the network.
Modularity is defined as the fraction of links that connect nodes
in the same community minus the fraction expected if
the partitioning and the degree sequence of the network remains fixed, 
but the links are randomly distributed~\cite{Qdef}.
Formally, for a network partitioning that assigns each node $i$ to
 one member of a set of communities, the modularity $Q$ is
\begin{equation}
\label{def.mod}
    Q=\frac{1}{2m}\sum_{i,j} B_{ij} \delta_{C(i),C(j)}
\end{equation}
where $B_{ij}=A_{ij}-k_i k_j/(2m)$ are the elements of the ``modularity matrix''
and $C(i)$($C(j)$) is the community to which node $i$($j$) belongs. Here $m$ is the total
number of links in the network, $k_i$($k_j$) is the degree of node $i$($j$), $A_{ij}$ are
the elements of the adjacency matrix, and $\delta$ is the Kronecker delta function.
The larger $Q_{\max}$, the maximum value of $Q$ for all network partitionings,
is for a network the more modular the network is.
The largest possible value of $Q_{\max}$ is one.

Unfortunately, finding the network partitioning that maximizes $Q$ is
known to be
an NP-hard problem and, thus, is computationally challenging~\cite{Nphard}.
In order to solve this problem, we use the
leading eigenvalue method combined with final tuning~\cite{FinalTuning}.
Final tuning improves
the approximate solution given by the leading eigenvalue method by
removing constraints that bias the results. For widely studied
example networks with up to a few thousand
nodes, the size of the genetic network of {\em E.\ coli} used in our analysis, combining final tuning
with the leading eigenvalue method has been demonstrated to produce network
partitionings with the largest $Q_{\max}$ of any known method~\cite{FinalTuning}. 

\subsection*{Creating artificial noisy datasets}

To explore the effects of experimental noise we found the community structure in artificial datasets created to mimic the actual data with various levels of experimental noise. To generate these datasets, we first considered a restricted set of the actual data consisting of the 152 experiments that were repeated at least three times in the ${\rm M^{3D}}$ database. For each of the 152 experiments we calculated the mean $m(X)$ and standard error $\sigma(X)$ of the expression level of each gene $X$. Assuming a normal distribution of error, we then generated artificial data for an ÒartificialÓ experiment by randomly choosing a value for the expression of each gene $X$ from a Gaussian distribution with mean $m(X)$ and standard deviation $c\:\sigma(X)$, where $c$ is a positive constant. The amount of noise in the artificial data can be adjusted by varying $c$ with $c = 0$ recreating the original
data set. Artificial data sets were generated at values of $c$ ranging from 0 to 4. For each value of $c$, ensembles of 20 different artificial data sets were constructed and then analyzed.

\subsection*{Statistical analysis of ensembles of network partitionings}

As noted above, many community detection algorithms, including the one we use, are stochastic in nature and 
can give diverse partitionings that maximize Q between different runs. We account for this by studying statistical properties of the ensemble of partitionings that result from repeated application 
of the community detection algorithm. In particular, we study the correlations of the partitionings in the ensemble and produced matrix correlation plots
that indicate the fraction of the pairs of partitionings for which pairs of genes are found to be in the same community. This ensemble analysis 
provides an understanding of the robustness of the community structure found. At the same time, it also provides information about the strength of the modular relationship between pairs of genes. 
 Note that this ensemble analysis method, unlike usual modularity maximizing methods of
community detection, allows for individual genes to be associated with more than one community. This is similar 
to information that can be obtained in, for example, clique~\cite{overlapping} and core-periphery~\cite{19426456} 
community detection methods.

\begin{figure}[ht]
\begin{center}
\includegraphics[width = 5.5in]{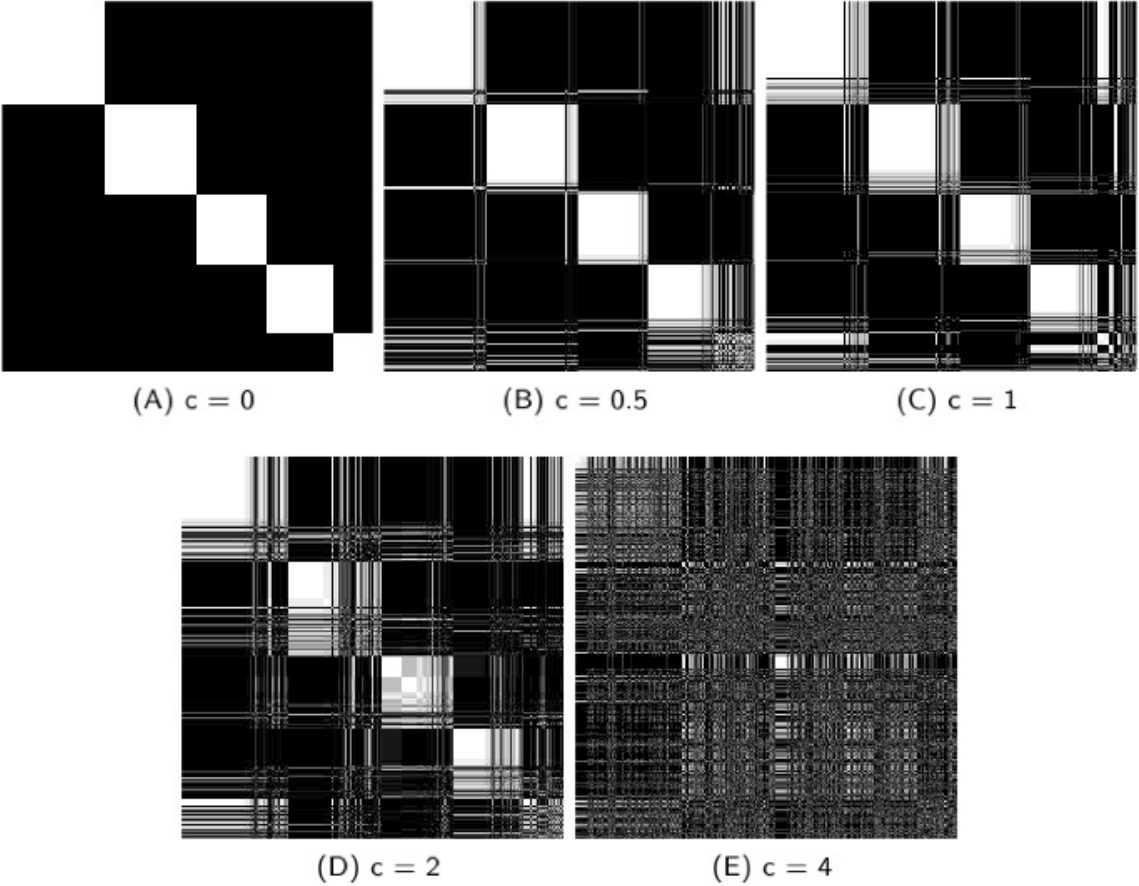}
\caption{
{\bf Change in core community structure as noise is increased from $c = 0$ to $c = 4$.} 
The
grey scale value of each element indicates the fraction of times the
two genes occurred in the same community over replicate community partitionings. If the element is white (black)
the two genes were always (never) found in the same community.
At each noise value there are clearly white diagonal blocks indicating sets of genes that are always found in the same community, which we refer to as core communities.
Note that, the five core communities at $c = 0$ (Fig.~\ref{core_communities}A ) are in the same order 
in Fig.~\ref{core_communities}:B, C, D, and E.
Within each of the five core communities of Fig.~\ref{core_communities}A , the node order is allowed to change in Fig.~\ref{core_communities}:B, C, D, and E in order
to display the largest subcommunity first. For each panel, he list of of the order of genes and the core community they belong to is given in Dataset S5 and Dataset S6, respectively.
A full size version with each pixel representing a distinct pair of genes is included in Fig. S2.}
\label{core_communities}
\end{center}
\end{figure}

The gray scale plots of Fig.~\ref{core_communities} are the matrix correlation plots for the statistical ensemble analysis of the 200 partitionings 
 constructed from the artificial noisy data for each noise value c.  
The grey scale of each matrix element in the plots corresponds
to the fraction of pairs of partitionings in which the corresponding pairs of genes are found to be in the same community. 
Note that the order of genes used in a matrix correlation plot is arbitrary. However, by judiciously choosing an ordering, modular relationships
become more apparent. 
The order of genes in Fig.~\ref{core_communities}A
is such that all of the genes in the largest core community are arbitrarily 
listed first, followed by a similar list of the genes in the second, third, fourth, and fifth largest core communities.  Note that when $c=0$ all genes are in one of the five core communities and therefore this list contains all genes. 
In Fig.~\ref{core_communities}:B, C, D, and E, the genes in each of the 5 core communities at $c=0$ have been reordered, 
but the order of the genes with respect to these core communities has been preserved. That is, in each of these subfigures, 
all genes that are in the $i^{th}$ largest core community at $c=0$ are always listed before any genes in the $j^{th}$ largest core community at $c=0$ if $i < j.$ In each subfigure, the genes within a $c=0$ core community have been reordered such that the subset of those genes that comprise the largest core community at the $c$ value corresponding to the subgraph are listed first, followed by those in the next largest such core community, etc. Until all genes within the $c=0$ core community has been listed. Note that, some genes may be isolated in their own core community with this method. 
  The list of the order of genes, for each subfigure, is given in Dataset S5.

The multicolor matrix correlation plot of Fig.~\ref{ori_c2_c3} simultaneously shows the statistical correlations in the modular relationships between pairs of genes, in the full dataset, at supercritical, critical, and subcritical threshold values.
 First, single color, blue red and green, matrix correlation plots  corresponding to  $f_{\min}$ values of 2, 4, and 6, respectively, are created. 
 The genes in each of these single color correlation plots are then simultaneously reordered as follows. 
 First, the genes were ordered so that all of the genes in the same community at $f_{\min} = 2$ are listed together, according to the size of the community, beginning with the largest and ending with the smallest. Next, the genes in each of those communities are reordered such that the subset of those genes that comprise the largest community at $f_{\min} = 4$ are listed first, followed by those in the next largest such community, etc. Until all genes within the $f_{\min} = 2$  community have been listed. Then each of the genes within a $f_{\min} = 4$ core community that are within an $f_{\min} = 2$ community are again reordered.  The genes in each of those communities are reordered such that the subset of those genes that comprise the largest core community at $f_{\min} = 6$  are listed first, followed by those in the next largest such community, etc. Until all genes within the $f_{\min} = 4$ core community that are within an $f_{\min} = 2$ community have been listed. The resulting ordering of genes is given in the 
 supplemental material. Finally, the three single color correlation plots are combined into the multicolor plot shown in Fig.~\ref{ori_c2_c3}, where
 each matrix element of the resulting plot has an RGB color that simultaneously indicates its correlations in the modular structure at each of the three
 $f_{\min}$ values. The list of the order of genes is given in Dataset S2.

\subsection*{Hypergeometric tests}

In order to establish the biological relevance of the functional communities found with our methods, 
we compare those functional communities to terms in the gene ontology
, using 
a hypergeometric test with Benjamini-Hochberg correction. 
The hypergeometric test calculates the probability that a community of size
$n$ has $k$ genes in common with a GO term of size $m$ in a network with
$N$ total genes. For random groupings this probability is
\begin{equation}
\label{hypergeo}
P
=
\frac{
\left(
\begin{array}{c}
m \\ k
\end{array}
\right)
\left(
\begin{array}{c}
N-m \\ n-k
\end{array}
\right)
}
{
\left(
\begin{array}{c}
N \\ n
\end{array}
\right)
}.
\end{equation} 
If a community and a GO term are found to have an overlap that is unlikely
to occur by chance (a low P value) then their
relationship is likely to be relevant. Note that a low P value can occur if the number of genes in common, k, is either greater than or
less than expected by chance. For a hypergeometric distribution the expected number of matches is given by $mn/N$.
We have reported only the ``positive'' enrichments for which  $k > mn/N$ as relevant.

To control for false discoveries due to multiple comparisons, we correct the $P$ values obtained using Eq.~\ref{hypergeo}
with the Benjamini-Hochberg (BH) procedure~\cite{BH}. We implement the BH procedure
as follows. For a given core community, the $P$ values obtained by comparing
it to the $M$ GO terms are ordered in a list such that they are increasing,
$P_1 \leq P_2 \leq \ldots \leq P_M$.
The corrected $P$ values are then taken to be $MP_r/r$, where $r$ is the rank, or position
on the ordered list, of the $P$ value. Then, as is commonly accepted, we judge
the relationship between a community and a GO term to be relevant if their
corrected $P$ value is less than 0.05.

To account for the resolution limit of modularity optimization~\cite{resolution}, only core communities
 of size 10 or larger are tested for biological relevance. The members of a GO term are 
 restricted to the genes included in our data set.

\section*{Acknowledgments}

We are grateful to G\'abor Bal\'azsi, Bogdan Danila, Dan Graur, and Kre\v simir Josi\'c for helpful discussions throughout this project. 

\bibliography{plos}

\newpage

\section*{Figures and Tables}

\begin{table}[ht]
\caption{
\bf{The 25 most relevant relationships found for $f_{\min} = 4$ without noise. }}
\begin{tabular}{c c r r r p{5.4cm}}
\hline \hline \\[-2ex]

P value & GO term num & Com size & GO size &
In common &
Description
\\ \hline

8.41e-42    &   9288        & 72     &  24          & 24    & bacterial-type flagellum  \\
9.57e-39    &   6826        & 53     &  37          & 25    & iron ion transport  \\
8.22e-38    &   1539        & 72     &  28          & 24    & ciliary or flagellar motility  \\
3.67e-35    &   6412        & 826    &  101         & 79    & translation  \\  
6.51e-34    &   3735        & 826    &  56          & 54    & structural constituent of ribosome  \\
3.08e-31    &   3723        & 826    &  105         & 77    & RNA binding  \\
1.73e-29    &   6935        & 72     &  22          & 19    & chemotaxis  \\
4.30e-29    &   3774        & 72     &  17          & 17    & motor activity  \\
5.38e-29    &   9425        & 72     &  17          & 17    & bacterial-type flagellum basal body  \\
2.06e-25    &  19861        & 72     &  15          & 15    & flagellum  \\
5.61e-25    &   5506        & 53     &  210         & 31    & iron ion binding  \\
3.72e-24    &  19843        & 826    &  42          & 40    & rRNA binding  \\
6.98e-23    &   6811        & 53     &  79          & 22    & ion transport  \\
6.99e-22    &  30529        & 826    &  36          & 35    & ribonucleoprotein complex  \\
1.72e-21    &   5840        & 826    &  38          & 36    & ribosome  \\
6.62e-21    &   8652        & 247    &  62          & 32    & cellular amino acid biosynthetic process  \\
4.11e-17    &   5506        & 139    &  210         & 39    & iron ion binding  \\
6.66e-16    &   9055        & 139    &  116         & 29    & electron carrier activity  \\
7.30e-15    &  51539        & 139    &  98          & 26    & 4 iron, 4 sulfur cluster binding  \\
8.22e-15    &  15453        & 300    &  15          & 15    & oxidoreduction-driven active transmembrane transporter activity light-driven active transmembrane transporter activity  \\
1.85e-13    &   6865        & 247    &  70          & 27    & amino acid transport  \\
6.13e-13    &  45272        & 300    &  13          & 13    & plasma membrane respiratory chain complex I  \\
9.19e-13    &  30964        & 300    &  13          & 13    & NADH dehydrogenase complex  \\
1.97e-12    &   9060        & 300    &  21          & 16    & aerobic respiration  \\
2.15e-12	 & 	5515 	& 826 	& 875 	& 251 & protein binding calmodulin binding \\
 \hline

\end{tabular}
\begin{flushleft}
The ``P value'' or random probability, calculated with a hypergeometric test with Benjamini-Hochberg correction, of the common occurrence, or overlap, of genes in an inferred community and in a GO term for the 25 most statistically relevant relationships are listed. Also listed are the ``GO term num'' that distinguishes the GO term and its ``Description'' in the GO database, the number of genes in the GO term ``GO size'', the number of genes in the inferred community ``Com size'', and the number of genes they have in common ``In common.'' 
The complete set of the 239 relevant relationships found 
for $f_{\min} = 4$, as well as the relevant relationships found for $f_{\min} = 2$ and $6$, are given in Dataset S7.
\end{flushleft}
\label{compare}
 \end{table}

\begin{center}
\begin{table}[ht]
   \caption{
   \bf{Genes in the community at  $f_{\min} = 6$  that enriches GO:3735 structural constituent of ribosome}}
   \label{OVERLAPf6}
   \begin{tabular}{  | p{8cm} | p{8cm} |}
    \hline
   Genes in the GO Term & Genes not in GO Term \\ \hline

 rplA, rplB, rplC, rplD, rplE, rplF, rplI, rplJ, rplK, rplL, rplM, rplN, rplO, rplP, rplQ, rplR, rplS, rplU, rplV, rplW, rplX, rplY,  & cdsA, cmk, dnaG, dusB, efp, fis, fusA, gidB, gmk, infB, ispU, lpxB, mnmG, mrdA, murA, nusA, nusG, obgE, parE,  \\ 
  rpmA, rpmB, rpmC, rpmD, rpmE, rpmG, rpmH, rpmJ, rpsA, rpsB, rpsC, rpsD, rpsE, rpsF, rpsG, rpsH, rpsI, rpsJ, rpsK,  &  ppa, prfC, priB, pyrH, queA, rbfA, rho, rimM, rlmN, rnhB, rnpA, rpoA, rpoZ, secE, secG, secY, speA, speB, tff, tig, \\ 
     rpsL, rpsM, rpsN, rpsO, rpsP, rpsQ, rpsR, rpsS, rpsT, rpsU, & trmA, trmD, trmI, truB, truC, tsf, typA, yadB, yggN, ygiQ, yhbC, yhbE, yhbY, yidC, yidD, yqcC  \\
   
    \hline
    \end{tabular}
 
   \end{table}
\end{center}


 \begin{figure}
\begin{center}
  \caption*{{\bf Dataset S1. List of links in the {\em E.\ coli} CLR network at $f_{\min} = 4$ and $6$}
The CLR algorithm is used to infer direct and indirect regulatory interactions between {\em E. coli}  genes on the basis of the similarity of their expression response in 466 experiments. A matrix of relatedness values is calculated and a network of regulatory interactions is inferred by placing links between
every pair of genes whose relatedness value exceeds some threshold, $f_{\min}$.  A list of the links for the network at 
$f_{\min} = 4$ and $f_{\min} = 6$ is provided. 
  Each link is given by listing a gene name, the gene's Blattner number, followed by the target gene name and Blattner number. 
  }
\end{center}
\end{figure}

 \begin{figure}
\begin{center}
  \caption*{
{\bf Dataset S2. List of the order of genes in the correlation matrix plot}
The multicolor matrix correlation plot simultaneously shows the statistical correlations in the modular relationships between pairs of genes, in the full dataset, at supercritical, critical, and subcritical threshold values.  Each matrix element of the resulting plot has an RGB color that simultaneously indicates its correlations in the modular structure at each of the three
 $f_{\min}$ values. A list of the order of genes is given, by listing each gene name and Blattner number.}
\end{center}
\end{figure}

 \begin{figure}
\begin{center}
  \caption*{
{\bf Dataset S3. List of core community membership}
At a particular $f_{\min}$ value, which defines a unique network,  the community detection algorithm was run 10 times, generating a correlation matrix where each element represents the proportion of times gene $X$ and gene $Y$ are found in the same community. 
Sets of genes that are always found in the same community is defined as a ``core community".  
 For each $f_{\min}$ value, 2, 4, and 6, the gene name, Blattner number and core community number is given. 
}
\end{center}
\end{figure}

 \begin{figure}
\begin{center}
  \caption*{
{\bf Dataset S4. The hierarchical organization of the {\em E.\ coli} network}
The relationship between communities at different $f_{\min}$ values indicates that the structure of the  {\em E. coli} network is largely hierarchical. A hierarchical structure is revealed when a community breaks up into subcommunities as $f_{\min}$  increases. Thus, if the  {\em E. coli}  regulatory network was completely hierarchical, one would see only block diagonal elements consisting of large blue blocks that break up into purple then white sub-blocks as $f_{\min}$ is increased. 
The hierarchical organization of genes is given, where the blue, purple, and white module membership of each gene is listed. The blue membership is listed first, numbered 1 through 6. The purple membership is listed next with the format x.y, where x is the blue membership and y is the purple membership. The purple membership is listed in the order a,b,c,d,....,z, aa, ab, .... Finally, the white membership is listed with format x.y.z, where x is the blue membership, y is the purple membership, and z is the white membership. The white membership is listed in numerical order. 
}
\end{center}
\end{figure}

 \begin{figure}
 \begin{center}
  \caption*{
{\bf Dataset S5. List of the order of genes in each noise correlation matrix plot}
The gray scale plots of Fig.~\ref{core_communities} are the matrix correlation plots for the statistical ensemble analysis of the 200 partitionings 
 constructed from the artificial noisy data for each noise value c.  
The grey scale of each matrix element in the plots corresponds
to the fraction of pairs of partitionings in which the corresponding pairs of genes are found to be in the same community. 
For each noise value $c = 0, 0.5, 1, 2$ and $4$, each ordered gene name and Blattner number is listed. 
}
\end{center}
\end{figure}

 \begin{figure}
\begin{center}
  \caption*{
{\bf Dataset S6. List of noise core community membership}
For each noisy data set, the CLR algorithm is used to infer a regulation network at an
$f_{\min}$ value of 2, and the community structure is
determined with the methods described above.
For each dataset, 10 different community partitionings are obtained, giving a total of 200 partititonings for each value of $c$.
Sets of genes that are always founds in the same community are defined as ``core communities".
For each noise value $c = 0, 0.5, 1, 2$ and $4$, a gene name and Blattner number, followed by its core community number is listed. 
}
\end{center}
\end{figure}

 \begin{figure}
\begin{center}
  \caption*{
{\bf Dataset S7. List of GO term enrichments}
 To determine whether the inferred community structure groups genes with similar biological functions, a test to determine whether the identified communities significantly enrich 
for any of the gene ontology (GO) terms identified in {\em E.\ coli} is performed. 
At each $f_{\min}$ value, each significant enrichment is listed by giving it's corresponding {\em p}-value, GO term number, core community number, community size, GO term size,
the number of genes in common, and the biological description of the GO term.  
}
\end{center}
\end{figure}

 \begin{figure}
 \begin{center}
  \caption*{
{\bf Dataset S8. List of GO terms}
 To determine whether the inferred community structure groups genes with similar biological functions, a test to determine whether the identified communities significantly enrich 
for any of the gene ontology (GO) terms identified in {\em E.\ coli} is performed. 
To test for enrichment, genes were removed from each GO term that were not included in our dataset. For each resulting GO term,  
the gene name and Blattner number, followed by its GO term number is listed.  
 }
\end{center}
\end{figure}

 \begin{figure}
\begin{center}
  \caption*{
{\bf Dataset S9. List of GO term enrichments at each noise value}
 To determine the effect of noise on GO term enrichment, at each noise value $c$, a test to determine whether the identified communities significantly enrich 
for any of the gene ontology (GO) terms identified in {\em E.\ coli} is performed. 
At each $c$ value, each significant enrichment is listed by giving it's corresponding p-value, GO term number, core community number, community size, GO term size,
the number of genes in common, and the biological description of the GO term.  
}
\end{center}
\end{figure}

 \cleardoublepage

 \begin{figure}
\begin{center}
  \caption*{{\bf Figure S1.  Correlation Matrix }
Correlation matrix showing community structure found in the {\em E. coli} network with relatedness threshold values $f_{min} = 2, 4,$ and $6.$
 Genes are ordered in the same sequence
along the x and y axes beginning in the upper left corner, and this ordering is the same for all three relatedness values (gene order is given in SI).
The matrix element
in the position $(X,Y)$ is colored blue, red, or green if genes $X$ and $Y$ are
in the same community at threshold values 2, 4, or 6, respectively.
The density of the color indicates the strength of the correlation in the partitionings of the pair of genes. For example, considering the correlation between a pair of genes in the 10 replicate partitionings performed on the $f_{min} = 4$ network, dark and light red indicates that the pair of genes 
are always and rarely found to be in the same community, respectively. 
The red, green and blue colors corresponding to $f_{min} = 2$, 4 and 6 thresholds, respectively, are combined to indicate the correlations of each pair of genes at all three threshold values.  
Thus, the color of the matrix element
in the position $(X,Y)$ is white if genes $X$ and $Y$ are
in the same community at all three threshold values.
It is purple (yellow) if the two genes are
in the same community at thresholds 2 and 4 (4 and 6), but not at threshold 6 (2) and it is black if the two genes are not in the same community at any of the three
threshold values. A list of the order of genes is given in Dataset S2.
}
\end{center}
\end{figure}

 \begin{figure}
 \begin{center}
  \caption*{{\bf Figure S2.  Noise Correlation Matrices  }
 Change in core community structure as noise is increased from $c = 0$ to $c = 4$.
The grey scale value of each element indicates the fraction of times the
two genes occurred in the same community over replicate community partitionings. If the element is white (black)
the two genes were always (never) found in the same community.
At each noise value there are clearly white diagonal blocks indicating sets of genes that are always found in the same community, which we refer to as core communities.
Note that, the five core communities at $c = 0$ (Fig.~\ref{core_communities}A ) are in the same order 
in Fig.~\ref{core_communities}:B, C, D, and E.
Within each of the five core communities of Fig.~\ref{core_communities}A , the node order is allowed to change in Fig.~\ref{core_communities}:B, C, D, and E in order
to display the largest subcommunity first. For each panel, he list of of the order of genes and the core community they belong to is given in Dataset S5 and Dataset S6, respectively.
}
\end{center}
\end{figure}

 \begin{figure}
\begin{center}
  \caption*{{\bf Figure S3.  Operon by Community }
Fraction of {\em E. coli} operons that are retained whole in a single community. 
The fraction of 544 operons (comprising 2172 genes) identified in the {\em E. coli} genome where all genes in the operon were assigned to the same final tuning community was determined at  $f_{\min} = 2, 4$ and $6$ (indicated by arrows). These actual values were compared to 1000 random distributions of the same set of genes to empty community sets of the same size and number as were present in the final tuning partitionings (histograms). In all cases, actual operon retention proportions were much greater than in any of the 1000 randomly distributed sets, indicating that they were very unlikely to occur by chance and therefore that the final tuning community partitionings effectively group genes in the same operon to the same community.
}
\end{center}
\end{figure}

 \begin{figure}
\begin{center}
  \caption*{{\bf Figure S4. Regulatory links from {\em flhDC} and {\em fliA} in the $f_{\min} = 4$ community that significantly enriches for flagellum associated genes}
 Genes are organized into operons as annotated by RegulonDB. Black, blue and red lines indicate regulatory interactions that are annotated in RegulonDB, inferred in the CLR network or both, respectively. For simplicity, only links from FlhDC and to targets of these links from {\em fliA} are shown. Many of the interactions that are found in the CLR network are not present in RegulonDB (blue lines). These interactions are candidates for indicating unrecognized regulatory interactions between FlhDC and the target genes. However, in most cases these interactions can be explained through the action of FlhDC on the sigma factor encoded by {\em fliA} (thick red line), which does directly affect all but one of the target genes. This point underlines the difference between the CLR network, which includes direct and indirect regulatory interactions, and the direct transcriptional network as annotated in RegulonDB. Note the CLR connection between FlhDC and the target gene {\em ymdA} cannot be explained through any known indirect interaction and is, therefore, a candidate for representing a new direct interaction. 
}
\end{center}
\end{figure}

 \begin{figure}
\begin{center}
  \caption*{{\bf Figure S5. Core community hierarchy}
 An alternative view of the core community hierarchy where each core community is represented by a node. The node label x.y indicates the $f_{\min}$ level, x, and core community number, y. The size of each node represents the number of genes in the community relative to communities at the same $f_{\min}$ level. The edge width and color value indicate the proportion of the ``daughter" community deriving from the connected ``parent'' community. For example, If all of the genes in a ``daughter" community are from one ``parent" community then there is one edge that is dark blue and thick. The nodes have been arranged to display the hierarchy of the network. 
}
\end{center}
\end{figure}

\begin{figure}
\begin{center}
  \caption*{{\bf Table S1.  The community with the most significant GO term enrichment at $f_{\min} = 4$ }
The community with the most significant GO term enrichment at $f_{\min} = 4$ contains 72 genes, including all 24 genes in the GO term for bacterial-type flagellum. The remaining 48 genes in this community are implicated as having some relevance for the development, function or control of the  {\em E. coli} flagellum.}
\end{center}
\end{figure}


\end{document}